\documentclass[11pt]{amsart}

\usepackage[hmargin=0.8in,height=8.6in]{geometry}
\usepackage{amssymb,amsthm}
\usepackage{delarray,verbatim}
\usepackage{natbib}

\usepackage{ifpdf}
\ifpdf
\usepackage[pdftex]{graphicx}
\else
\usepackage[dvips]{graphicx}
\fi

\usepackage{ifpdf}
\usepackage{color}
\definecolor{webgreen}{rgb}{0,.5,0}
\definecolor{webbrown}{rgb}{.8,0,0}
\definecolor{emphcolor}{rgb}{0.95,0.95,0.95}

\usepackage{hyperref}
\hypersetup{%
          colorlinks=true,
          linkcolor=webbrown,
          filecolor=webbrown,
          citecolor=webgreen,
          breaklinks=true}
\ifpdf
\hypersetup{pdftex,
            pdfstartview=FitH, 
            bookmarksopen=true,
            bookmarksnumbered=true
}
\else
\hypersetup{dvips}
\fi

\numberwithin{equation}{section} 

\newtheorem {thm}{Theorem}[section]

\newtheorem {lemm}{Lemma}[section]

\theoremstyle{remark}
\newtheorem {rem}{Remark}[section]

\newcommand{\la}{\lambda}

\renewcommand{\P}{\mathbb P}

\newcommand{\E}{\mathbb E}

\renewcommand{\bar}{\overline}

\title[]{Maximizing Utility of Consumption Subject to a Constraint on the Probability of Lifetime Ruin}

\author[]{Erhan Bayraktar}

\address[E. Bayraktar]{Department of
  Mathematics, University of Michigan, Ann Arbor, MI 48109}
\email{erhan@umich.edu}
\thanks{E. Bayraktar thanks the National Science Foundation for financial support. }

\author[]{Virginia R. Young}
\address[V. R. Young]{Department of
  Mathematics, University of Michigan, Ann Arbor, MI 48109}
\email{vryoung@umich.edu}
\thanks{V. R. Young thanks the Cecil J. and Ethel M. Nesbitt Professorship for financial support.
}


\begin{document}

\begin{abstract}
In this note, we explicitly solve the problem of maximizing utility of consumption (until the minimum of bankruptcy and the time of death) with a constraint on the probability of lifetime ruin, which can be interpreted as a risk measure on the whole path of the wealth process.
\end{abstract}

\maketitle

\section{Introduction} \label{sec:intro}

In our past work, we determined the optimal investment strategy of an individual who targets a  given rate of consumption and who seeks to minimize the probability of going bankrupt before she dies, also known as lifetime ruin.  For an economic justification of minimizing lifetime probability of ruin as an investment criterion see, for example,  \cite{MR2324574}. \cite{MR2100933} considered this problem when the individual continuously consumes either a constant (real) dollar amount or a constant proportion of wealth. \cite{MR2324574} introduced borrowing constraints.  \cite{MR2295829} established when the two problems of minimizing a function of lifetime minimum wealth and of maximizing utility of lifetime consumption result in the same optimal investment strategy on a given open interval in wealth space. On the other hand, \cite{bayyou08} considered the case for which the consumption is stochastic and is correlated to the wealth.  By using a convex duality relationship between the stopper and controller problems, we showed that the minimal probability of lifetime ruin in this set up, whose stochastic representation does not have a classical form as the utility maximization problem does, is the unique classical solution of its Hamilton-Jacobi-Bellman (HJB) equation, which is a non-linear boundary-value problem. 


When we presented our work several of our colleagues suggested that we consider the probability of lifetime ruin as a risk constraint.  Portfolio optimization problems with risk constraints on terminal wealth were considered by \cite{basshp01} and \cite{MR2332260}, for example.  Their solution can be obtained by pathwise maximization since the problem can formulated in terms of choosing the terminal optimal wealth.  However, the probability of lifetime ruin is a risk constraint on the entire path of the wealth process (the constraint depends on whether the lifetime minimum reaches the bankruptcy level before the time of death), and such an approach is not possible.  Risk constraints on the entire path of the process recently were also considered by \cite{MR2211713}.  By using the results of \cite{MR844005}, we will show that optimal strategies investment and consumption strategies exist and can be numerically obtained using a bisection search.

The rest of the paper is organized as follows: In Section 2.1, we provide a precise statement of the problem.  In Section 2.2, we give sufficient conditions for optimality and relate the problem to an unconstrained optimization problem (Lemma 2.1), which was analyzed by \cite{MR844005}.  In Section 3.1, we summarize the results of  \cite{MR844005} that we need to prove our main result.  In Section 3.2, we prove our main result, Theorem 3.1, with the help of the auxiliary results Lemmas 3.3-3.5. These results point out that a bisection search can be carried out to determine an optimal policy (see Remark~\ref{rem:bisection}).

\section{Maximizing Utility of Consumption Subject to a Probability of Ruin Constraint}

In Section 2.1, we present the financial market and define the problem of maximizing expected utility of lifetime consumption subject to a constraint on the probability of lifetime ruin.   In Section 2.2, we show that the method of Lagrange multipliers allows us to apply the work of \cite{MR844005} to solve this problem.

\subsection{Statement of the Problem}

In this section, we first present the financial ingredients that make up the agent's wealth, namely, consumption, a riskless asset, and a risky asset.  We, then, define the problem of maximizing utility of consumption subject to a constraint on the ruin probability.  We assume that the individual invests in a riskless asset whose price $S^0_t$ at time $t$ follows the process $dS^0_t = r \, S^0_t \, dt$, for some fixed rate of interest $r > 0$.  Also, the individual invests in a risky asset whose price at time $t$, $S_t$, follows geometric Brownian motion given by
\begin{equation}
dS_t = \mu \, S_t \, dt + \sigma \, S_t \, dZ_t,
\end{equation}
in which $\mu > r$, $\sigma > 0$, and $Z$ is a standard Brownian motion with respect to a filtered probability space $(\Omega, \mathcal{F}, \P, \{\mathcal{F}_t\}_{t \geq 0})$.  Let $X_t$ denote the wealth at time $t$ of the investor.  We allow him to specify a consumption process $\{ c_t \}_{t \ge 0}$ and an investment strategy $\{ \pi_t \}_{t \ge 0}$, in which $c_t$ denotes the rate of consumption and $\pi_t$ denotes the dollar amount invested in the risky asset at time $t$.   (We will consider {\it admissible} consumption and investment pairs $\{c_t,\pi_t\}_{t \geq 0}$ such that $\{c_t\}_{t \geq 0}$ is a non-negative process adapted to $\{\mathcal{F}\}_{t \geq 0}$ and satisfies $\int_0^t c_s ds<\infty$ almost surely for all $t \geq 0$, and such that $\{\pi_t\}_{t \geq 0}$ is adapted to $\{\mathcal{F}\}_{t \geq 0}$ and  satisfies $\int_0^t \pi_s^2 \, ds < \infty$ almost surely for all $t \ge 0$.)  The remaining wealth, namely $X_t - \pi_t$, is invested in the riskless asset.  Thus, wealth follows the process
\begin{equation}
dX_t = [rX_t + (\mu - r) \pi_t  - c_t] \, dt + \sigma \, \pi_t \, dZ_t,  \quad X_0=x,
\end{equation}

We assume that the investor chooses admissible consumption and investment strategies to maximize his expected utility of consumption before he dies or before he bankrupts, whichever occurs first.  Let $U$ denote a strictly increasing, strictly concave utility function on $(0, \infty)$ whose first three derivatives are continuous; we extend $U$ to $[0, \infty)$ by defining $U(0) = \lim_{c \downarrow 0} U(c)$; similarly, for $U'(0)$.   Let $\tau_d$ denote the random time of death of the investor.  We assume that $\tau_d$ is exponentially distributed with parameter $\beta >0$ (that is, with expected time of death equal to $1/\beta$); this parameter is also known as the {\it hazard rate} of the investor.  We assume that $\tau_d$ is independent of the Brownian motion $Z$ driving the price process of the risky asset.

The investor is subject to a constraint on the probability of bankruptcy, or ruin, before he dies.  Let $\tau_0$ denote the first time that wealth equals $0$; that is, $\tau_0 = \inf \{ t \ge 0: X_t = 0 \}$.  $\tau_0$ is the time of ruin or bankruptcy.  For notational completeness, we should write $\tau^{\{ c_t, \pi_t \}}_0$ to indicate the dependence of $\tau_0$ on the consumption and investment strategies.  However, for simplicity, we will write $\tau_0$, if the consumption and investment strategies are understood by the context.

Thus, the investor chooses admissible $\{ c_t \}_{t \ge 0}$ and $\{ \pi_t \}_{t \ge 0}$ to maximize
\begin{equation}\label{eq:W}
W_{ \{c_t, \pi_t \}}(x) := \E^x \left[ \int_0^{\tau_0 \wedge \tau_d} U(c_t) \, dt \right],
\end{equation}
subject to the constraint
\begin{equation}\label{eq:prob-ruin}
\psi(x) := \P^x (\tau_0 < \tau_d) \le \varphi(x).  
\end{equation}
Here, $\varphi(x)$ is a given threshold of tolerance for the probability of lifetime ruin.  We assume that $\varphi(x)$ is less than or equal to the probability of ruin when the individual maximizes \eqref{eq:W} {\it without} any constraint on the probability of ruin.  $\E^x$ and $\P^x$ denote the expectation and probability, respectively, conditional on $X_0 = x$.  

\begin{rem}
 Note that because $\tau_d \sim Exp(\beta)$ is independent of the Brownian motion driving the price process of the risky asset, we can express $W_{\{c_t, \pi_t \}}$ as follows.
\begin{equation}
\begin{split}
W_{\{ c_t, \pi_t \}}(x) &= \E^x \left[ \int_0^{\tau_d} U(c_t) \, {\bf 1}_{\{ t \le \tau_0 \}} \, dt \right] =  \E^x \left[ \int_0^\infty \beta e^{-\beta s} \, \int_0^s  U(c_t) \, {\bf 1}_{\{ t \le \tau_0 \}} \, dt \, ds \right] \\
&=  \E^x \left[ \int_0^\infty  U(c_t) \, {\bf 1}_{\{ t \le \tau_0 \}} \,  \int_t^\infty \beta e^{-\beta s}  \, ds \, dt \right]  \\
&= \E^x \left[ \int_0^\infty  e^{- \beta t} \, U(c_t) \, {\bf 1}_{\{ t \le \tau_0 \}} \, dt \right]  = \E^x \left[ \int_0^{\tau_0}  e^{- \beta t} \, U(c_t) \, dt \right].
\end{split}
\end{equation}
Similarly, we can rewrite the ruin probability in \eqref{eq:prob-ruin} as follows:
\begin{equation}\label{eq:prob-ruin2}
\begin{split}
 \psi(x) &=  \E^x \left[  \int_0^\infty  {\bf 1}_{\{ \tau_0 \le t \}} \, \beta e^{-\beta t} \, dt \right] =  \E^x \left[  \int_{\tau_0}^\infty  {\bf 1}_{\{ \tau_0 < \infty \}} \, \beta e^{-\beta t} \, dt \right] \cr
&= \E^x \left[  e^{-\beta \tau_0}  {\bf 1}_{\{ \tau_0 < \infty \}} \right] = \E^x \left[  e^{-\beta \tau_0} \right].
\end{split}
\end{equation}

\end{rem}

\begin{rem}
To ensure that maximizing \eqref{eq:W} subject to \eqref{eq:prob-ruin} results in a finite value function, we make the following assumption concerning the utility function $U$.  Define the positive constant $\gamma$ by
\begin{equation}
\gamma = {1 \over 2} \left( {\mu - r \over \sigma} \right)^2,
\end{equation}
and let $\la_- < -1$ and $\la_+ > 0$ denote the solutions of
\begin{equation}
\gamma \la^2 - (r - \beta - \gamma) \la - r = 0. 
\end{equation}
We assume that
\begin{equation}
\int_c^\infty {d \theta \over (U'(\theta))^{\la_-}} < \infty,
\end{equation}
for all $c > 0$.
\end{rem}

\subsection{Sufficient Conditions for Optimality}

For any real number $P$ and admissible strategy $\{ c_t, \pi_t \}_{t \ge 0}$,  define 
\begin{equation}\label{eq:defn-V}
V_{\{c_t,\pi_t\}}(x;P)=\E^x \left[\int_0^{\tau_0}e^{-\beta t }U(c_t)dt+P e^{-\beta \tau_0}\right].
\end{equation}

\begin{lemm}\label{eq:lagrange}
For a fixed $x>0$, assume that there exists an admissible strategy $\{ c^*_t, \pi^*_t \}_{t \ge 0}$ such that
\begin{equation}
\P^x \left(\tau_0^{\{c^*_t, \pi^*_t \}} \leq \tau_d \right)=\varphi(x), 
\end{equation}
and assume that there exists a constant $P \le 0$ such that $\{ c^*_t, \pi^*_t \}_{t \ge 0}$ solves 
\begin{equation}\label{eq:defn-Vstar}
V^*(x;P):=\sup_{ \{ c_t, \pi_t \}} V_{\{c_t, \pi_t \}}(x;P)=V_{\{c_t^*,\pi_t^*\}}(x;P).
\end{equation}
Then, for any admissible strategy $\{ c_t, \pi_t \}_{t \ge 0}$ satisfying $\P^x \left(\tau_0^{\{c_t, \pi_t \}} \leq \tau_d \right) \leq \varphi(x)$, we have that
\begin{equation}
\E^x \left[\int_0^{\tau_0^{ \{c_t, \pi_t \}}} e^{-\beta t } \, U(c_t) \, dt\right] \leq \E^x \left[\int_0^{\tau_0^{ \{c^*_t, \pi^*_t \}}} e^{-\beta t } \, U(c_t) \, dt\right].
\end{equation}
\end{lemm}

\proof{By assumption
\begin{equation}\label{eq:defn-optimality}
\E^x \left[\int_0^{\tau_0^{\{c_t, \pi_t \}}} e^{-\beta t } \, U(c_t) \, dt+P e^{-\beta \tau_0^{\{c_t, \pi_t \}}}\right] \leq \E^x \left[\int_0^{\tau_0^{\{c^*_t, \pi^*_t \}}} e^{-\beta t } \, U(c_t) \, dt+P e^{-\beta \tau_0^{\{c^*_t, \pi^*_t \}}}\right].
\end{equation}

On the other hand, again from our assumption, $\E^x \left[e^{-\beta \tau_0^{\{c^*_t, \pi^*_t \}}}\right]=\varphi(x) \geq \E^x \left[e^{-\beta \tau_0^{ \{c_t, \pi_t \}}}\right]$ for any admissible $\{c_t, \pi_t \}_{t \ge 0}$.  Using this last inequality in \eqref{eq:defn-optimality} yields the result.

\medskip

Note that the expressions for $V_{\{c_t, \pi_t \}}$ in \eqref{eq:defn-V}  and $V^*$ in \eqref{eq:defn-Vstar} are identical to the corresponding expressions in (1.6) and (2.7) in \cite{MR844005}.  Thanks to the results of that paper, we will be able to identify $P$, $\{c^*_t \}_{t \ge 0}$, and $\{ \pi^*_t \}_{t \ge 0}$ satisfying the assumptions of  Lemma~\ref{eq:lagrange}.

\section{Main Result}

In this section, we will first describe how to obtain $V^*$ for a given real number $P$ using the results presented in \cite{MR844005}. Then, we will give our main result and show the existence of $P$, $\{ c^*_t \}_{t \ge 0}$, and $\{\pi^*_t \}_{t \ge 0}$.  We will also describe how these can be numerically computed efficiently. 

\subsection{Solution of the Unconstrained Optimization Problem in \eqref{eq:defn-Vstar}}\label{sec:KLSS}

Karatzas et al.\ (1986; Theorem 4.1 and Remark 4.2) prove the following verification theorem concerning $V^*$:

\begin{lemm}For $P$ finite, suppose $V: (0, \infty) \to (P, \infty)$ is a $C^2$ function satisfying the Hamilton-Jacobi-Bellman (HJB) equation
\begin{equation}\label{eq:verf}
\beta V(x) = rx V'(x) + \max_{c \ge 0} \left( U(c) - c V'(x) \right) + \max_\pi \left( (\mu - r) \pi V'(x) + {1 \over 2} \sigma^2 \pi^2 V''(x) \right). 
\end{equation}
\item{$(a)$} If $U(0)$ is finite, then $V(x) \ge V^*(x)$ for $x > 0$.
\item{$(b)$} If $U(0) = -\infty$ and if $\E^x \left[ \int_0^{\tau_0} e^{-\beta t} \max(0, U(c_t)) \, dt \right] < \infty$ for all admissible strategies, then $V(x) \ge V^*(x)$ for $x > 0$.
\end{lemm}

When the non-negativity constraint on consumption is not active, then \cite{MR844005} (Sections 6 through 12) use consumption as an intermediate variable to solve the HJB equation \eqref{eq:verf}.  When the non-negativity constraint on consumption is active, then in Section 13, they use $y = (V^*)'(x)$ as an intermediate, or dual, variable.  We shall follow the latter approach in both cases and thereby unify our presentation of the results of \cite{MR844005}.

Define the function $I: (0, U'(0)] \to [0, \infty)$ to be the inverse of $U'$.  Extend $I$ to $(0, \infty)$ by setting $I \equiv 0$ on $[U'(0), \infty)$.  If $V$ is $C^2$ and strictly concave, then the HJB equation (3.1) becomes
\begin{equation}
\beta V(x) = \left[ rx - I(V'(x)) \right] V'(x) + U(I(V'(x))) - \gamma {(V'(x))^2 \over V''(x)}, \quad x > 0.  
\end{equation}

Let $\rho_- < 0$ and $\rho_+ > 1$ denote the solutions of
\begin{equation}\label{eq:qdrteqn}
\gamma \rho^2 - (r - \beta + \gamma) \rho - r = 0.
\end{equation}
Note that $\rho_\pm = 1 + \la_\pm$.  For $a \geq 0$, $B \leq 0$, and $A \leq 0$, define the following functions for $0 < y < U'(a)$.  
\begin{equation}
\begin{split}
\mathcal{X}(y; a, B) &= B y^{\la_+} +  {I(y) \over r}  -{1 \over \gamma(\la_+ - \la_-)} \left\{ {y^{\la_+} \over \la_+} \int_a^{I(y)} {d \theta \over (U'(\theta))^{\la_+}} +  {y^{\la_-} \over \la_-} \int_{I(y)}^\infty {d \theta \over (U'(\theta))^{\la_-}} \right\},
\end{split}
\end{equation}
and
\begin{equation}
\begin{split}
\mathcal{J}(y; a, A) &= A y^{\rho_+} +  {U(I(y)) \over \beta} -{1 \over \gamma(\la_+ - \la_-)} \left\{ {y^{\rho_+} \over \rho_+} \int_a^{I(y)} {d \theta \over (U'(\theta))^{\la_+}} +  {y^{\rho_-} \over \rho_-} \int_{I(y)}^\infty {d \theta \over (U'(\theta))^{\la_-}} \right\}.
\end{split}
\end{equation}
Now, $\mathcal{X}$ is strictly decreasing, and $\lim_{y \downarrow 0} \mathcal{X}(y; a, B) = \infty$.  Define $\mathcal{X}(U'(a); a, B) = \lim_{y \uparrow U'(a)} \mathcal{X}(y; a, B)$.  Thus, $\mathcal{X}(\cdot; a, B)$ maps $(0, U'(a)]$ onto $[\mathcal{X}(U'(a); a, B), \infty)$, and its inverse function $\mathcal{Y}(\cdot; a, B)$ is $C^2$, strictly decreasing, and maps $[\mathcal{X}(U'(a); a, B), \infty)$ onto $(0, U'(a)]$.

When $U(0)$ and $U'(0)$ are finite, and when
\begin{equation}
{1 \over \beta} U(0) \le P < P^* := {1 \over \beta} U(0) - {(U'(0))^{\rho_-} \over \beta \la_-} \int_0^\infty {d \theta \over (U'(\theta))^{\la_-}}, 
\end{equation}
we set $a = 0$ and extend the definition of $\mathcal{X}$ and $\mathcal{J}$ to all $y > 0$.  In this case, $\mathcal{X}$ is strictly decreasing for all $y > 0$, and $\lim_{y \to \infty} \mathcal{X}(y; 0, B) = \lim_{y \to \infty} B y^{\la_+}$.  If $P = U(0)/\beta$, then set $B = 0$, so $\mathcal{X}(\cdot; 0, B)$ maps $(0, \infty)$ onto $(0, \infty)$ and has surjective inverse $\mathcal{Y}(\cdot; 0, B): (0, \infty) \to (0, \infty)$.  If $U(0)/\beta < P < P^*$, then set
\begin{equation}\label{eq:B}
B(P) = - {\beta \over \gamma(\la_+ - \la_-) [\bar y(P)]^{\rho_+}} \left( P - {1 \over \beta} U(0) \right),
\end{equation}
in which $\bar y(P)$ is defined by
\begin{equation}\label{eq:bar-y}
 [\bar{y}(P)]^{\rho_-} = - \beta \la_- \left( P - {1 \over \beta} U(0) \right) \left[ \int_0^\infty {d \theta \over (U'(\theta))^{\la_-}} \right]^{-1}. 
\end{equation}
It follows that $I(\bar y) = 0$ and $\mathcal{X}(\bar y; 0, B) = 0$.  We consider $\mathcal{X}(\cdot; 0, B)$ restricted to $(0, \bar y]$, which has range $[0, \infty)$ and surjective inverse $\mathcal{Y}(\cdot; 0, B): [0, \infty) \to (0, \bar y]$.

Define $V$ by
\begin{equation}\label{eq:y-to--val}
V(x; a, B) = {\mathcal{J}} \left( \mathcal{Y}(x; a, B ); a, {\la_+ \over \rho_+} B \right),
\end{equation}
with the domain of $V$ given by the domain of $\mathcal{Y}$.  Define candidate optimal consumption and investment strategies in feedback form by
\begin{equation}\label{eq:opt-st}
c_t = I(V'(X_t; a, B)), \qquad \qquad \pi_t = - \, {\mu - r \over \sigma^2} \, {V'(X_t; a, B) \over V''(X_t; a, B)}.  
\end{equation}

After these preliminaries, we are ready to give $V^*(x;P)$ in terms of $V(x;a,B)$ the optimal consumption and investment strategies by specifying the values of $a$ and $B$ in \eqref{eq:y-to--val} and \eqref{eq:opt-st} case by case; see Table 1 on page 292 of \cite{MR844005}.  First, we dispense with the case for which $P \ge \lim_{c \to \infty} U(c)/\beta$.  In this case, it is optimal to consume all of one's wealth immediately to bankruptcy; no continuous optimal consumption strategy exists.

\begin{lemm}\label{lem:KLSS}
Depending on the utility function and its derivative at zero, the value function $V^*(x;P)$ can be computed in terms of $V(x;a,B)$ as follows:
\begin{enumerate}
\item When $P \le U(0)/\beta$,  $V^{*}(x;P)=V(x;0,0)$.  Note that $V(x;0,0)$ is independent of $P$.
\item When $U'(0) = \infty$ and $U(0)/\beta < P < \lim_{c \to \infty} U(c)/\beta$,  $V^*(x;P)=V(x,a(P),B(P))$, in which given $P$, $a(P)$ is the unique positive root of the strictly decreasing function
\begin{equation}\label{eq:F-func}
F(c)=\rho_+ P - {(U'(c))^{\rho_-} \over \gamma \la_- \rho_-} \int_c^\infty {d \theta \over (U'(\theta))^{\la_-}} - {\rho_+ \over \beta} U(c) + {\la_+ \over r} c U'(c), 
\end{equation}
and $B(P)$ is defined by
\begin{equation}\label{eq:B-2}
B(P)\cdot (U'(a))^{\la_+} + {a \over r} - {(U'(a))^{\la_-} \over \gamma \la_- (\la_+ - \la_-)} \int_a^\infty {d \theta \over (U'(\theta))^{\la_-}} = 0.  
\end{equation}

\item  When $U'(0)$ finite and $U(0)/\beta < P < P^*$, $V^{*}(x;P)=V(x;0,B(P))$, in which $B(P)$ is given by \eqref{eq:B}.

\item  $U'(0)$ finite and $P = P^*$,  $V^{*}(x;P)=V(x;0,B(P))$, in which $B(P)$ is given by \eqref{eq:B-2}.

\item $U'(0)$ finite and $P^* < P < \lim_{c \to \infty} U(c)/\beta$,  $V^*(x;P)=V(x,a(P),B(P))$, in which $a(P)$ and $B(P)$ are as in Case ii.
\end{enumerate}
The strategy 
\begin{equation}\label{eq:opt-st}
c^P_t = I((V^*)'(X_t; P)), \qquad \qquad \pi^P_t = - \, {\mu - r \over \sigma^2} \, {(V^*)'(X_t; P) \over (V^*)''(X_t; P)},  
\end{equation}
satisfies
\begin{equation}
V_{\{c_t^P,\pi_t^P\}}(x;P)=V^*(x;P).
\end{equation}
\end{lemm}

Except for Case iii, the non-negativity constraint on the rate of consumption is not active.  For Case iii, the optimal rate of consumption is $0$ for wealth between $0$ and $\bar x$ and positive for wealth greater than $\bar x$, in which $\bar x = {\mathcal{X}}(U'(0); 0, B(P))$ with $B(P)$ given by \eqref{eq:B}.

\subsection{Probability of Lifetime Ruin}

Define a process $\{ Y_t \}_{t \ge 0}$ by $Y_t := (V^*)'(X_t)$, in which $V^*$ is given in Lemma~\ref{lem:KLSS}.  \cite{MR844005} show that bankruptcy occurs when the process $\{ Y_t \}_{t \ge 0}$ hits $\bar y(P)$, in which $\bar y(P)$ is given by \eqref{eq:bar-y} for Case iii and equals $U'(a(P))$ for the remaining cases; see Remarks 7.1 and 13.3 in that article.  Note that $U'(a)$ might be infinite, in which case the probability of bankruptcy is $0$.

\cite{MR844005} show that the process $\{ Y_t \}_{t \ge 0}$ follows geometric Brownian motion.  Specifically,
\begin{equation}
dY_t = - (r - \beta) \, Y_t \, dt - {\mu - r \over \sigma} \, Y_t \, dZ_t. 
\end{equation}
Recall from \eqref{eq:prob-ruin2} that the probability of lifetime ruin equals $\psi(x) = \E^x(e^{-\beta \tau_0})$.  Thus, for $y(P)$ satisfying $\mathcal{X}(y;a(P),B(P))=x$ (or $y(P)=(V^*)'(x;P))$, the probability of lifetime ruin is  $ \psi(x)=\hat \psi(y(P))$ in which $\hat{\psi}(y)$ solves
\begin{equation}\label{eq:pde-prob-ruin}
\begin{split}
& \beta \, \hat \psi(y) = -(r - \beta) \, y \, \hat \psi'(y) + \gamma \, y^2 \, \hat \psi''(y), \quad 0 \le y \le \bar{y}(P); \cr
& \hat \psi(\bar{y}(P)) = 1.
\end{split}
\end{equation}
From \eqref{eq:pde-prob-ruin}, it is straightforward to prove the following result.

\lemm{The probability of lifetime ruin $\psi$ is given by $\psi(x) = \hat \psi(y)$, with $y$ satisfying $\mathcal{X}(y;a,B)=x$, in which $\hat \psi$ equals
\begin{equation}\label{eq:exp-exp-rpb-ruin}
\hat{\psi}(y;P) = \left( y(P)/ \bar{y}(P) \right)^{\rho_+}, \quad y \in [0, \bar{y}(P)].
\end{equation}
Recall that $\rho_+ > 1$ is the positive root of \eqref{eq:qdrteqn}.}

\rem \label{rem:probuinzero} $\bar y = \infty$ in Case i in Lemma~\ref{lem:KLSS}. When $P \leq U(0)/\beta$, then the probability of bankruptcy, $\P^x(\tau_0<\infty)$, is zero ( see Remarks 7.1 and 13.3 in \cite{MR844005}), and therefore the probability of lifetime ruin, $\P^x(\tau_0<\tau_d)$, is also equal to zero.

\medskip

Now, given a threshold $\varphi$ for the probability of lifetime ruin, we wish to determine a corresponding Lagrange multiplier $P$ in \eqref{eq:defn-V}.  Then, by the solution of \eqref{eq:defn-Vstar} given in Lemma~\ref{lem:KLSS}, we have the corresponding optimal investment and consumption strategies, as stated there.

First, we show that the probability of lifetime ruin in \eqref{eq:exp-exp-rpb-ruin} is increasing with respect to $P$.

\lemm \label{lem:increasing}{As a function of the penalty $P \in [U(0)/\beta, \lim_{c \to \infty} U(c)/\beta],$ the probability of lifetime ruin $\hat \psi$ is increasing.  Moreover, when $P = U(0)/\beta,$ the probability of lifetime ruin equals $0,$ and as $P$ approaches $\lim_{c \to \infty} U(c)/\beta$, the probability of lifetime ruin approaches $1$.}

\proof  {The limiting values of the probability with respect to $P$ follow from Remark~\ref{rem:probuinzero}  and the discussion preceding Lemma~\ref{lem:KLSS}.  Next, let $P_1 < P_2$; we wish to show that $\E^x(e^{-\beta \tau_0})$ under the optimal consumption and investment strategy ${\{c^{(1)}_t, \pi^{(1)}_t \}}$ corresponding to $P_1$ is less than or equal to $\E^x(e^{-\beta \tau_0})$ under the optimal consumption and investment strategy ${\{c^{(2)}_t, \pi^{(2)}_t \}}$ corresponding to $P_2$.  From the optimality of ${\{c^{(2)}_t, \pi^{(2)}_t \}}$ for $P_2$, we have
\begin{equation}
\E^x\left[ \int_0^{\tau_0^{\{c^{(1)}_t, \pi^{(1)}_t \}}}  e^{- \beta t} \, U(c^{(1)}_t) \, dt + e^{-\beta \tau_0^{\{c^{(1)}_t, \pi^{(1)}_t \}}} P_2 \right]  \le \E^x\left[ \int_0^{\tau_0^{\{c^{(2)}_t, \pi^{(2)}_t \}}}  e^{- \beta t} \, U(c^{(2)}_t) \, dt + e^{-\beta \tau_0^{\{c^{(2)}_t, \pi^{(2)}_t \}} }P_2 \right], 
\end{equation}
and we can rewrite this inequality as
\begin{equation}\label{eq:inc-P1}
\begin{split}
 \E^x &\left[ \int_0^{\tau_0^{\{c^{(1)}_t, \pi^{(1)}_t \}}}  e^{- \beta t} \, U(c^{(1)}_t) \, dt + e^{-\beta \tau_0^{\{c^{(1)}_t, \pi^{(1)}_t \}}} P_1 + e^{-\beta \tau_0^{\{c^{(1)}_t, \pi^{(1)}_t \}}} (P_2 - P_1) \right]  \cr
 &\le \E^x\left[ \int_0^{\tau_0^{\{c^{(2)}_t, \pi^{(2)}_t \}}}  e^{- \beta t} \, U(c^{(2)}_t) \, dt  +   e^{-\beta \tau_0^{\{c^{(2)}_t, \pi^{(2)}_t \}}} P_1 + e^{-\beta \tau_0^{\{c^{(2)}_t, \pi^{(2)}_t \}}} (P_2 - P_1)\right].
\end{split}
\end{equation}
On the other hand, from the optimality of ${\{c^{(1)}_t, \pi^{(1)}_t \}}$ for $P_1$, we have
\begin{equation}\label{eq:inc-P2}
\E^x \left[ \int_0^{\tau_0^{\{c^{(2)}_t, \pi^{(2)}_t \}}}  e^{- \beta t} \, U(c^{(2)}_t) \, dt + e^{-\beta \tau_0^{\{c^{(2)}_t, \pi^{(2)}_t \}}} P_1 \right] \le\E^x\left[ \int_0^{\tau_0^{\{c^{(1)}_t, \pi^{(1)}_t \}}}  e^{- \beta t} \, U(c^{(1)}_t) \, dt + e^{-\beta \tau_0^{\{c^{(1)}_t, \pi^{(1)}_t \}}}P_1 \right] .
\end{equation}
From \eqref{eq:inc-P1} and \eqref{eq:inc-P2}, it follows that
\begin{equation}
 \E^x \left[e^{-\beta \tau_0^{\{c^{(1)}_t, \pi^{(1)}_t \}}}\right]  \le  \E^x\left[ e^{-\beta \tau_0^{\{c^{(2)}_t, \pi^{(2)}_t \}}}\right] ,
\end{equation}
which is what we wished to show. } 

\rem\label{rem:P0}{ Note that if $P = 0$, then the probability of lifetime ruin corresponding to $V^*$ in \eqref{eq:defn-Vstar} equals the probability of ruin when we maximize $W$ in \eqref{eq:W} without the constraint in \eqref{eq:prob-ruin}.  Therefore, $P = 0$ attains the maximum allowable value for the threshold $\varphi(x)$, namely, the probability of ruin when the individual maximizes $W$ without any constraint on the probability of ruin.}

\lemm\label{lem:C1}{The probability of lifetime ruin defined in \eqref{eq:exp-exp-rpb-ruin} is a continuously differentiable function of $P$.}

\proof{To show that $\hat{\psi}$ is a continuously differentiable implicit function of $P$, we will need to determine that (given $x$) $y$ and $\bar{y}$ are continuously differentiable functions of $P$. 
Note that $y$ is a solution $\mathcal{X}(y;a(P),B(P))=x$ (for proper values of $a$ and $B$, whose choice depends on the value function $U$ as described by Lemma~\ref{lem:KLSS}).  Since $\mathcal{X}(\cdot;a,B)$ is strictly decreasing and continuously differentiable, the implicit function theorem implies that $y$ is a continuously differentiable function of $P$.  On the other hand, $\bar{y}$ is either given by \eqref{eq:bar-y}, by $U'(a)$, in which $a$ is the unique root of the strictly decreasing function $F$ in \eqref{eq:F-func}, or by $U'(0)$.  One only needs to prove that $U'(a)$ is a continuously differentiable function of $P$. But, this result again follows from the implicit function theorem thanks to the fact that $F$ is strictly decreasing and differentiable.}

\thm \label{thm:main}{For a given wealth $x > 0$, there exists a constant $P \le 0$ such that the strategy $\{c^P_t, \pi^P_t \}_{t \ge 0}$, defined by \eqref{eq:opt-st}, maximizes expected utility of consumption \eqref{eq:W} subject to the probability of ruin constraint \eqref{eq:prob-ruin}.}

\proof{Let $U(0)/\beta \ge 0$.  Recall that if $P \le U(0)/\beta$ in \eqref{eq:defn-Vstar}, ruin is impossible under the optimal strategy.  Therefore, from Remark~\ref{rem:P0}, we can choose $P = 0$.

Now, let $U(0)/\beta < 0$.  Thanks to Lemmas~\ref{lem:increasing} and \ref{lem:C1} and to Remark~\ref{rem:P0}, for a fixed $x>0$, we can determine a $P \le 0$ satisfying $\P^x \left(\tau_0^{\{c^P_t, \pi^P\}} \leq \tau_d \right)=\varphi(x) \in [0,1]$. The result follows now from Lemmas  \ref{eq:lagrange} and \ref{lem:KLSS}.

\rem\label{rem:bisection}{It follow from  Lemmas~\ref{lem:increasing} and \ref{lem:C1} that for a given $x>0$, the constant $P$ in Theorem~\ref{thm:main} can be computed by bisection search using \eqref{eq:exp-exp-rpb-ruin} along with $\mathcal{X}(y;a,B)=x$ and the expression for $\bar{y}$ described in Lemma~\ref{lem:KLSS}.}

\bibliography{references}

\begin{thebibliography}{8}
\expandafter\ifx\csname natexlab\endcsname\relax\def\natexlab#1{#1}\fi
\expandafter\ifx\csname url\endcsname\relax
  \def\url#1{{\tt #1}}\fi

\bibitem[Basak and Shapiro(2001)]{basshp01}
S.~Basak and A.~Shapiro.
\newblock Value at risk based risk management: Optimal policies and asset
  prices.
\newblock {\em J. Business}, 78\penalty0 (3):\penalty0 1215--1266, 2001.

\bibitem[Bayraktar and Young(2007{\natexlab{a}})]{MR2295829}
E.~Bayraktar and V.~R. Young.
\newblock Correspondence between lifetime minimum wealth and utility of
  consumption.
\newblock {\em Finance Stoch.}, 11\penalty0 (2):\penalty0 213--236,
  2007{\natexlab{a}}.
\newblock ISSN 0949-2984.

\bibitem[Bayraktar and Young(2007{\natexlab{b}})]{MR2324574}
E.~Bayraktar and V.~R. Young.
\newblock Minimizing the probability of lifetime ruin under borrowing
  constraints.
\newblock {\em Insurance Math. Econom.}, 41\penalty0 (1):\penalty0 196--221,
  2007{\natexlab{b}}.
\newblock ISSN 0167-6687.

\bibitem[Bayraktar and Young(2008)]{bayyou08}
E.~Bayraktar and V.~R. Young.
\newblock Proving the regularity of the minimal probability of ruin via a game
  of stopping and control.
\newblock Technical report, University of Michigan, 2008.
\newblock URL \url{http://arxiv.org/abs/0704.2244v2}.

\bibitem[Boyle and Tian(2007)]{MR2332260}
P.~Boyle and W.~Tian.
\newblock Portfolio management with constraints.
\newblock {\em Math. Finance}, 17\penalty0 (3):\penalty0 319--343, 2007.
\newblock ISSN 0960-1627.

\bibitem[Cheridito et~al.(2005)Cheridito, Delbaen, and Kupper]{MR2211713}
P.~Cheridito, F.~Delbaen, and M.~Kupper.
\newblock Coherent and convex monetary risk measures for unbounded c\`adl\`ag
  processes.
\newblock {\em Finance Stoch.}, 9\penalty0 (3):\penalty0 369--387, 2005.
\newblock ISSN 0949-2984.

\bibitem[Karatzas et~al.(1986)Karatzas, Lehoczky, Sethi, and Shreve]{MR844005}
I.~Karatzas, J.~P. Lehoczky, S.~P. Sethi, and S.~E. Shreve.
\newblock Explicit solution of a general consumption/investment problem.
\newblock {\em Math. Oper. Res.}, 11\penalty0 (2):\penalty0 261--294, 1986.
\newblock ISSN 0364-765X.

\bibitem[Young(2004)]{MR2100933}
V.~R. Young.
\newblock Optimal investment strategy to minimize the probability of lifetime
  ruin.
\newblock {\em N. Am. Actuar. J.}, 8\penalty0 (4):\penalty0 105--126, 2004.
\newblock ISSN 1092-0277.

\end{thebibliography}
\bibliographystyle{abbrvnat}

\end{document}